\documentclass[9pt]{article}

\usepackage{amsmath,amssymb} 
 \usepackage[]{graphicx}
\graphicspath{{graphics/}} 
 \usepackage{epstopdf}
\usepackage{multirow}
\usepackage{color}
 \usepackage{wrapfig}
 \usepackage{ulem}
\usepackage{lscape}
\usepackage{fullpage}

\def\red{\textcolor{black}}

\def\eps{\varepsilon}
\newcommand{\fraz}{\displaystyle\frac}
\def\##1{{\bf #1}}
\def\=#1{\underline{\underline #1}}
\def\.{\mbox{ \tiny{$^\bullet$} }}
\def\ux{\#u_x}
\def\uy{\#u_y}
\def\uz{\#u_z}
\def\uprop{\#u_{prop}}

\def\ped0{_{\scriptscriptstyle 0}}
\def\k0{k\ped0}
\def\tq{\tilde{q}}
\def\vph{v_{ph}}
\def\propdist{\Delta_{prop}}
\def\fref#1{Figure~\ref{#1}}

\def\tref#1{Table~\ref{#1}}
\def\tond#1{\left(#1\right)}
\def\quadr#1{\left[#1\right]}
\def\graff#1{\left\{#1\right\}} 
\def\dblq#1{\textquotedblleft #1\textquotedblright}
\newcommand*\diff{\mathop{}\!\mathrm{d}}

\begin{document}
\begin{center}
 
\textbf{Compound Guided Waves That Mix Characteristics of Surface-Plasmon-Polariton, 
	Tamm, Dyakonov--Tamm, and Uller--Zenneck Waves}\\ \vspace{4mm}

{Francesco Chiadini},$^1$
{Vincenzo Fiumara},$^2$
{Antonio Scaglione},$^1$ and
{Akhlesh Lakhtakia}$^3$\\ \vspace{4mm}

$^1${Department of Industrial Engineering,
	University of Salerno, via Giovanni Paolo II, \\ 132 - Fisciano (SA), 
	84084, Italy}\\ \vspace{4mm}
$^2${School of Engineering, University of 
	Basilicata, Viale 
	dell'Ateneo Lucano 10, 85100 Potenza, Italy}\\ \vspace{4mm}
$^3${Department of Engineering Science and Mechanics, Pennsylvania State University,
	\\ University Park, PA 16802--6812,
	USA} \vspace{4mm}

\end{center}

\noindent \textbf{Abstract.}
Solutions of the boundary-value problem for electromagnetic   waves guided by a layer of a homogeneous and isotropic (metal or dielectric) material sandwiched between a structurally chiral material (SCM) and a periodically multi-layered isotropic dielectric (PMLID) material were  numerically obtained and analyzed. If the sandwiched layer is sufficiently thick, the two bimaterial interfaces decouple from each other, and each interface may guide one or more electromagnetic surface waves (ESWs) by itself. Depending on the constitution of the two materials that partner to form an interface, the ESWs can be classified as surface-plasmon-polarition (SPP) waves, Tamm waves, Dyakonov--Tamm waves, or Uller--Zenneck waves. When the sandwiched layer is sufficiently thin, the ESWs for single bimaterial interfaces coalesce to form compound guided waves (CGWs). The phase speeds, propagation distances, and spatial profiles of the electromagnetic fields of   CGWs are different from those of the ESWs. The energy of a CGW is distributed in both the SCM and the PMLID material, if the sandwiched layer is sufficiently thin. Some CGWs require the sandwiched layer to have a minimum thickness. Indeed, the coupling between the two  faces of the sandwiched layer is affected by  the ratio of the thickness of the sandwiched layer to the skin depth in that material and  the rates at which the fields of the ESWs guided individually by the two interfaces decay away from their respective guiding interfaces.\\

\section{Introduction} \label{sec:intro}

Several different kinds of electromagnetic surface waves (ESWs) exist. Perhaps the first ever analytical treatment was of an ESW guided by the planar interface of air and seawater, published in his doctoral dissertation by  Uller in 1903 \cite{Uller}. As Wait \cite{Wait} recounted in a review of ground-wave propagation, that was also the time when  Tesla \cite{Tesla}
conjectured that radio waves are guided by the air/earth interface, a proposition analyzed by Zenneck \cite{Zenneck} in 1907 and by Sommerfeld \cite{Sommerfeld1909,Sommerfeld1920,Sommerfeld1926} in several papers thereafter. The
ESWs bound to the planar interface of two homogeneous dielectric materials of which one is dissipative are called Zenneck waves, but should be more properly called Uller--Zenneck waves \cite{UZ-Faryad}.

ESWs called surface-plasmon-polariton (SPP) waves are guided by the planar interface of two isotropic and homogeneous materials, one of which has a relative permittivity with negative real part (usually, a metal)
and the other whose relative permittivity has positive real part (usually, a dielectric material). 
SPP waves became a topic of research after  the energy losses of
electrons impinging on a metal film were explained in 1957 in terms of electronic-plasma oscillations
occurring at the film's surfaces \cite{Ritchie}. But SPP waves cannot be excited by shining light directly at either a dielectric film lying atop a metal film or a metal film lying atop a dielectric film. The reflectance of a parallel-polarized plane
wave from a thin aluminum film deposited on the hypotenuse
of a right-angled glass prism was shown in 1959
to exhibit a sharp dip as the
angle of incidence exceeded the critical angle for a planar glass/air interface \cite{Turbadar}. The
reflectance dip was related to the excitation of an SPP wave  in 1968 \cite{Otto, KR}, and an explosive expansion of research on applications of SPP waves started thereafter \cite{Pitarke,AZLreview,ZZX}.
The excitation of a multiplicity of SPP waves at a specific frequency is possible if the dielectric partnering material is periodically non-homogeneous  (on the order of wavelength) in the direction normal to the interface~\cite{AkhBook,LiuJNP}. In this case, even though all SPP waves are excited at the same frequency, they can differ for polarization state, phase speed, attenuation rate, and field profile; furthermore, the dielectric partnering material can even be anisotropic \cite{Sprokel,AkhBook}.

Most SPP waves do
not propagate long distances along the interface plane, because metals are dissipative.
If both partnering materials are weakly dissipative dielectric materials,
the propagation distance would be enhanced.
Indeed, the Tamm wave was predicted in 1977 to be guided
by the interface of two isotropic dielectric materials, at
least one of which was periodically non-homogeneous in
the direction perpendicular to the interface \cite{YYH}.  The
experimental observation of Tamm waves followed in 1978
\cite{YYC}, and their application to optical sensing has been subsequently
demonstrated
\cite{Shinn,Konopsky}. 

A few years later, Marchevski\u{i} \textit{et al.} \cite{MSS}
and Dyakonov \cite{Dyakonov} predicted an ESW guided
by the interface of two homogeneous dielectric materials,
at least one of which is anisotropic. Experimental
verification came only in 2009, when Takayama \textit{et al.}
were able to excite a Dyakonov wave guided by the interface
of a liquid and a biaxial dielectric crystal \cite{Takayama1}.
But the Dyakonov wave  propagates in very narrow ranges of directions ($\lesssim1$~deg
out of a maximum of $360$~deg)
in the interface plane \cite{Takayama2}, and is difficult to detect.
The insertion of a 10--20-nm thick dielectric sheet between an anisotropic
and an isotropic material can enhance the range of directions
of propagation \cite{Takayama3}.

In 2007 Lakhtakia and Polo \cite{LP2007} proposed an
ESW that is guided by the interface
of two dielectric materials, one of which is isotropic
and \red{homogeneous,} and the other is both anisotropic and
periodically non-homogeneous in the direction perpendicular
to the interface plane. Combining the attributes of both
Tamm and Dyakonov waves, this surface wave was named
a Dyakonov--Tamm  wave. The angular sectors permitting
the propagation of Dyakonov--Tamm  waves are so large as to often cover the
entire 360 deg available \cite{AkhBook}. Dyakonov--Tamm waves 
are also expected to be useful for optical sensing \cite{LFJNP}.

Two SPP waves can be \textit{compounded} when  a
metal layer of thickness $L$ is
sandwiched between two half spaces filled with  isotropic, homogeneous  dielectric materials
\cite{Economou,Wendler,Yang}. 
But $L$ must not be significantly larger than the skin depth $\delta$ in the
metal, or the coupling between the two dielectric/metal interfaces  vanishes and each of the
two interfaces by itself  guides an SPP wave that will not coalesce
\red{with the other SPP wave}.  The compound guided wave (CGW) can propagate
over longer distances than an SPP wave guided by a single metal/dielectric
interface.  Recently, we examined the  CGWs guided by a thin metal layer
sandwiched between a  homogeneous isotropic dielectric (HID) material and either a  periodically multi-layered isotropic dielectric (PMLID) material \cite{NoiJNP15} or a structurally chiral material (SCM) \cite{NoiOC15}, all dielectric non-metallic materials being assumed to have negligible dissipation. We found that a multiplicity of CGWs with plasmonic and polaritonic constituents can propagate bound to  both metal/dielectric interfaces  with energy distributed in both the dielectric materials if the metal layer is sufficiently thin. Long-range CGWs are also guided by a thin metal layer
inserted in a periodically non-homogeneous anisotropic material such that
the direction of non-homogeneity is the same as the thickness direction of the metal
film   \cite{FL-PSSRRL} with non-homogeneity on the wavelength scale; furthermore, CGWs resembling Dyakonov--Tamm waves
in their characteristics are \red{guided,} if the
sandwiched metal  is replaced by an isotropic dielectric material  \cite{FL-PRA2011}.
Finally,  CGWs that fuse together the characteristics of SPP waves and Dyakonov waves have been theoretically shown to excited via   a uniaxial dielectric layer inserted between a HID material and a metal \cite{Takayama4}.
Thus, planar multi-interface structures offer the possibility of CGWs that mix the characteristics
of two (or more)  ESWs. 

With our interest lying in the compounding of ESWs that require at least one of the partnering materials to be periodically non-homogeneous normal to the direction of propagation,
in the present \red{paper}, we address and solve the  boundary-value problem  of the propagation of a CGW guided by a layer of a homogeneous and isotropic (metal or dielectric) material sandwiched between an SCM and a PMLID material. Thereby, compounding of 
the characteristics of SPP waves, Tamm waves,  Dyakonov--Tamm waves, 
and Uller--Zenneck
waves is investigated. Let us note that
the analyzed structures can be manufactured using techniques routinely used to fabricate optical thin films, especially multilayer thin films~\cite{Baum,HWbook} and sculptured thin films \cite{AkhBook2}, and liquid crystals \cite{deG}.     

The plan of this paper is as follows: Sec.~\ref{sec:matmeth} provides in brief
 the boundary-value problem for CGWs guided by a planar {\sf A}/{\sf B}/{\sf C}
 structure and introduces the materials  chosen as  {\sf A}, {\sf B},
 and {\sf C} for the numerical results presented in Secs.~\ref{sec:oneinterface}  and
 \ref{sec:twointerfaces}. Section~\ref{sec:oneinterface} provides some results for ESWs guided by  the {\sf A}/{\sf B},   {\sf B}/{\sf C}, and  {\sf A}/{\sf C} bimaterial interfaces individually.
 Section~\ref{sec:twointerfaces} is devoted to  numerical results for CGWs
 guided by the   {\sf A}/{\sf B}/{\sf C} structure and comparison with the results
 presented in Sec.~\ref{sec:oneinterface}.
Concluding remarks follow in Sec.~\ref{sec:cr}.

An $\exp\tond{-i\omega t}$  dependence on time $t$ is implicit, with $\omega$ denoting the angular frequency and $i=\sqrt{-1}$. The free-space wavenumber, the free-space wavelength, and the intrinsic impedance of free space are denoted by $\k0=\sqrt{\eps\ped0 \mu\ped0}$, $\lambda_0=2\pi/\k0$, and $\eta_0=\sqrt{\mu\ped0/\eps\ped0}$, respectively, with $\eps\ped0$ and $\mu\ped0$ being the permeability and permittivity of free space. The speed of light in vacuum is denoted by $c\ped0=\sqrt{\eps\ped0\mu\ped0}$. Vectors are in boldface; dyadics are underlined twice; and  Cartesian unit vectors are identified as $\ux$, $\uy$, and $\uz$.

\section{Theoretical Preliminaries and Materials}\label{sec:matmeth}
\begin{figure}[h!]
	\centering
	\includegraphics[width=0.8\linewidth]{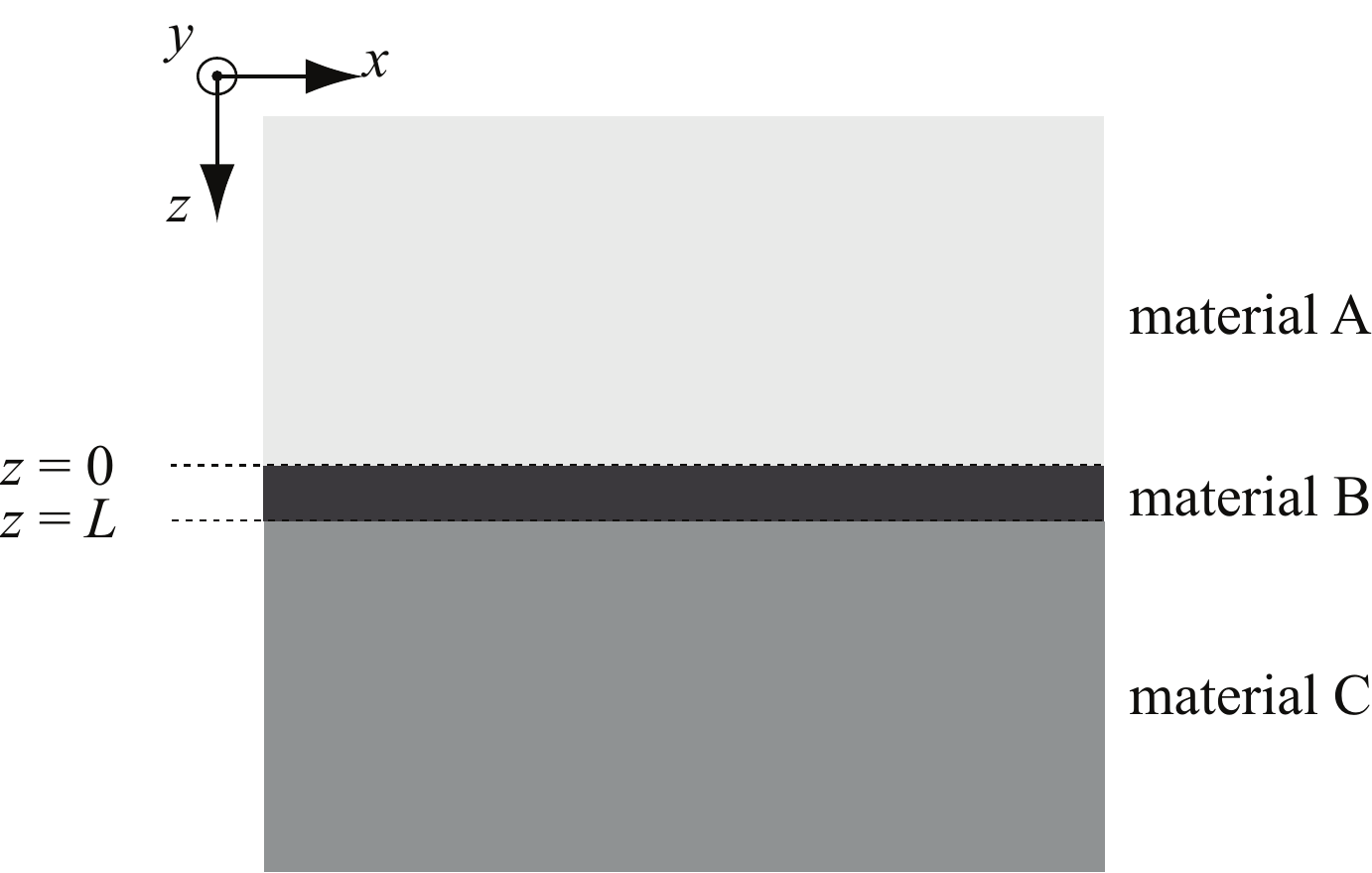}
	\caption{Schematic of the boundary-value problem to examine the propagation of CGWs.}
	\label{fig:schematic}
\end{figure}

A schematic of the boundary-value problem for the propagation of a CGW is provided in \fref{fig:schematic}. A slab of material {\sf B}  is interposed between the half~space $z<0$ occupied by material {\sf A} and the  half~space $z>L$ occupied by a material {\sf C}.
Given the wide variety of compound CGWs possible, we chose  {\sf A} to be a PMLID material, {\sf B} to be a homogeneous and isotropic material, and {\sf C} to be an SCM.

 The PMLID material {\sf A} has  a period  $\Lambda= 675$~nm with a unit cell consisting of  $N=9$  
equal-thickness \red{nondissipative} dielectric layers of different silicon oxynitrides, each fabricated by plasma-enhanced chemical
 vapor deposition of a specific composition of silane, ammonia,
 and nitrous oxide \cite{PMLIDmet}. The layers contain
SiO$_2$ and SiN$_x$ in the  ratios shown in \tref{tab:PMLID}. The layer labeled
$j=1$ in the first unit cell of the PMLID material is the one closest to material {\sf B}.

\begin{table}[h!]
\centering
\caption{\bf Composition and relative permittivity $ \eps_j $ at $ \lambda_0=633 $~nm of the j$^{th}$  layer in the unit cell of  the
chosen  PMLID material \cite{PMLIDmet}.
\label{tab:PMLID}
}\vspace{0.4cm}
\begin{tabular}{c|ccc}

$ j $ & SiO$_2\%$& SiN$_x\%$ & $\eps_j $\\ \hline
$1 $ & $0$ & $100$ &$ 3.9357$\\ $2 $ & $32$ & $68$ & $ 3.3033$\\
$ 3$ & $40$ & $60$ &$ 3.1515 $ \\ $ 4 $& $49$ & $51$ &$ 3.0056$  \\
$ 5 $  & $62$ &$38$ &$ 2.7526 $\\ $ 6 $ & $72$ & $28$ & $ 2.6143 $ \\ 
$ 7 $& $82$ & $18$ & $ 2.4475 $  \\ $ 8 $ & $90$ & $10$ & $  2.3161 $ \\ $ 9 $ & $100$ & $0$ & $ 2.1837 $  \\
\hline
\end{tabular}
\end{table}
 
For all calculations reported here,  material {\sf B} can be
 either (i) a metal, specifically silver with relative permittivity $\eps_{Ag}=-14.461+i 1.1936$;
 or (ii) a  dissipative HID with  $\eps_{dHID}=14.461+i 1.1936$ which was used 
 as a \red{nonplasmonic} analog of silver; or
 (iii) a practically \red{nondissipative} HID, specifically a glass with relative permittivity $\eps_{glass}=2.56$.
 
 Material {\sf C} was chosen to be
 an SCM---specifically, a structurally right-handed chiral sculptured thin film of patinal titanium oxide
 with permittivity dyadic given at $ \lambda\ped0=633$~nm by~\cite{NoiOC15,HWH} 
 \begin{equation}
 \begin{aligned}
 &\=\eps_{SCM}\left(z\right) =   \eps\ped0\=S_z\left(\frac{2\pi z}{P}\right)
 \. \=S_y\left(37.6745^\circ \right)\\
 & \.\left( 2.13952\uz \uz + 3.66907\ux\ux 
 + 2.82571 \uy \uy\right) \\
 &\.\=S_y^{-1}\left(37.6745^\circ \right)\. \=S_z^{-1}\left(\frac{2\pi z}{P}\right)\,,
 \end{aligned}\label{epschiral}
 \end{equation}
 with
 \begin{eqnarray}
 \nonumber
 &&\=S_z\tond{\zeta}= \uz\uz + \left(\ux \ux  + \uy \uy\right)\cos \zeta  \\ 
 &&\qquad+\left(\uy \ux- \ux \uy\right)\sin \zeta\,,
 \label{Sz}
 \\
 \nonumber
 &&\=S_{y}\left(\chi \right)=  \uy \uy + \left(\ux \ux+ \uz \uz\right)\cos  \chi   \\ 
 &&\qquad+ \left(\uz \ux- \ux \uz\right)\sin \chi \,,
 \label{Sy}
 \end{eqnarray}
 and $P=270$~nm as the period along the $z$ axis.

 We consider the CGW to be  propagating parallel to the unit vector $\uprop=\ux\cos\psi+\uy\sin\psi$, $\psi\in[0^\circ,360^\circ)$, in the  $xy$ plane  and decaying far away from the layer of material {\sf B}. 
With  $q$ as  the complex-valued wavenumber of the CGW,
the electric and magnetic field phasors   can be represented everywhere  by
\begin{equation}
\left.\begin{array}{l}
 \#E(\#r)= \#e(z) \exp\left({iq}\uprop\.\#r\right)\\
\#H(\#r)= \#h(z) \exp\left({iq}\uprop\.\#r\right)
\end{array}\right\}\,.
\label{eq:EH1}
\end{equation}

 The axial field components $e_z(z)$ and $h_z(z)$ of $\#e(z)$
and $\#h(z)$, respectively, can be
expressed in terms of the column 4-vector \cite{AkhBook}
\begin{equation}
\quadr{\#f\tond{z}}=\begin{bmatrix}
e_x(z) & e_y(z) & h_x(z) & h_y(z) 
\end{bmatrix}^{T}
\end{equation}
that satisfies the matrix differential
equations
\begin{equation}
\fraz{\diff }{\diff z}\quadr{\#f\tond{z}}= 
i\quadr{\=P^{(\sigma)}\tond{z}}\.\quadr{\#f\tond{z}},\quad \sigma= \begin{cases}
{\sf A}, & z< 0\\
{\sf B}, & 0< z< L\\
{\sf C}, & z> L
\end{cases}
\label{eq:ODE}
\end{equation}
where the $4\times 4$ matrixes $\quadr{\=P^{(\sigma)}\tond{z}}$ can be  written for materials {\sf A}, {\sf B}, and {\sf C}.
Since  material {\sf B} is homogeneous, the matrix $\quadr{\=P^{({\sf B})}}$ is independent of $z$, and we obtain
\begin{equation}
\quadr{\#f\tond{L}}=\exp\left\{i
\quadr{\=P^{(B)}}L\right\}
\.
\quadr{\#f\tond{0}}
\label{eq:PB}
\end{equation}
after solving \eqref{eq:ODE}~\cite{AkhBook}.
By virtue of the periodicity of both materials {\sf A}, and {\sf C} along the $z$ axis, we have
\begin{eqnarray}
\quadr{\#f\tond{0}}=&\quadr{\=Q^{\sf \tond{A}}}
\.
\quadr{\#f\tond{-\Lambda}}\,
\label{eq:Q-}\\
\quadr{\#f\tond{L+P}}=&\quadr{\=Q^{\sf \tond{C}}}
\.
\quadr{\#f\tond{L}}\label{eq:Q+}
\end{eqnarray}
where each of the $4\times 4$ matrixes $\quadr{\=Q^{\sf \tond{A}}}$ and $\quadr{\=Q^{\sf \tond{C}}}$ characterizes the optical response of one period. Whereas $\quadr{\=Q^{\sf \tond{A}}}$
can be calculated as the cascade of the transfer matrixes of the layers constituting
one period of the PMLID material,  $\quadr{\=Q^{\sf \tond{C}}}$ can be obtained by a piecewise uniform approximation subdividing one period of the SCM in thin slices, parallel to the plane $ z=0 $, with equal thickness and spatially uniform dielectric properties~\cite{AkhBook,AkhBook2}.

 After the imposition of the standard boundary conditions across the interfaces $z=0$ and $z=L$, Eqs.~(\ref{eq:PB})--(\ref{eq:Q+}) lead to the matrix equation
\begin{equation}
\quadr{\=M\tond{\red{q}}}\cdot\quadr{\#S}=0\,.
\label{eq:M}
\end{equation}
The matrix $\quadr{\=M\tond{\red{q}}}$ is synthesized from the   $\quadr{\=Q^{\sf \tond{B}}}$ as well as from the eigenvectors of  $\quadr{\=Q^{\sf \tond{A}}}$ and  $\quadr{\=Q^{\sf \tond{C}}}$, as shown in detail
 elsewhere~\cite[Sec.~3.6]{AkhBook}.
The column 4-vector $\quadr{\#S}$ containins 4 unknown coefficients.

 Equation~(\ref{eq:M}) has a nontrivial solution only when the matrix
$\quadr{\=M\tond{\red{q}}}$ is singular. Therefore the dispersion equation $\det\quadr{\=M\tond{q}}=0$ has to be solved for $q$.

We numerically solved the dispersion equation  to obtain the normalized wavenumbers $\tq =q/k\ped0 $ of the CGWs. 
Knowing $q$,  we can calculate the phase speed    $ \vph=c\ped0/ {\rm Re}\left(\tq\right)$   and the propagation distance $\propdist=1/{\rm Im}(q)$ of the CGW in the direction parallel to $\uprop$, where $\propdist$ is the distance along the direction of propagation at which the wave amplitude reduces by a factor of \red{$\exp({-1})= 0.367$ and} the power density reduces by $\exp({-2}) = 0.135$.
The  main  characteristics  of the CGW can be illustrated by the  spatial distribution of the time-averaged Poynting vector
\begin{equation}
\#P(\#r) 
=\fraz{1}{2}{\rm Re}\left[\#e(z) \times \#h^\ast(z)\right]
\exp[-2{\rm Im}(q)\uprop\.\#r]\,,
\label{eq:Poynt}
\end{equation}
where the asterisk denotes the complex conjugate.
If  both materials {\sf A} and {\sf C} are  isotropic, the CGW is either $p$ or $s$
polarized, and $q$ as well as $\propdist$ are independent of $\psi$ (i.e., the direction of
propagation in the $xy$ plane). But, if at least one of those two materials is an SCM, then no polarization state can be assigned to the CGW \cite{AkhBook}. 

We focused on CGWs that fuse together at least two ESWs of the
following kinds: SPP waves, Tamm waves,  Dyakonov--Tamm waves, 
and Uller--Zenneck
waves. In order to present representative numerical results, we fixed 
$ \lambda_0=633 $~nm and $\psi=30$~deg,
 the dependence on $\psi$ of these CGWs being quite weak for the chosen SCM.
All calculations were  restricted to  $0<{\rm Re}(\tq) \leq 3$ to avoid computational
instabilities that emerged for ${\rm Re}(\tq) > 3$.

\section{ESWs guided by an {\sf A}/{\sf B},   {\sf B}/{\sf C}, or  {\sf A}/{\sf C} bimaterial interface}\label{sec:oneinterface}

In order to appreciate the characteristics of CGWs guided by two parallel bimaterial interfaces (as in
Fig.~\ref{fig:schematic}), it is best to briefly present results for simple ESWs guided by a single bimaterial interface.  \tref{tab:taxon} summarizes the type of ESWs guided by  the planar interface of two different partnering materials:  {\sf A}/{\sf B},   {\sf B}/{\sf C}, or  {\sf A}/{\sf C}. Their relevant characteristics and the differences between ESWs of different types are discussed in the following subsections.

\begin{table*}[h!]\setlength\tabcolsep{4pt}
 	\centering
 	\caption{\bf Types of ESWs and their characteristics for the choice of partnering materials $\sf A$ and $\sf B$.
 		\label{tab:taxon}}\vspace{4mm}
 	\begin{tabular}{llllll}
 		\textbf{ESW Type} & \multicolumn{2}{l}{\textbf{Partnering materials}}& \textbf{Data} &\textbf{Number} & \textbf{Characteristics}\\
 		\cline{2-3} & \textbf{A} &\textbf{B}& & \textbf{of modes} &\\ \hline
 		SPP & SCM & Metal &\tref{tab:qSCMmet} & $2$ &  No polarization state can be assigned\\  
 		 & PMLID material & Metal & \tref{tab:qPMLIDmet} & $7$ &  4 $p$- and 3 $s$-polarized, 2 high-phase speed solutions\\ \hline
 		Uller--Zenneck & SCM & Dissipative HID & \tref{tab:qSCM-HID} & $2$ &  No polarization state can be assigned\\ 
 		 & PMLID material &  Dissipative HID & \tref{tab:qPMLID-HID} & $6$ &  3 $p$- and 3 $s$-polarized, 2 high-phase speed solutions\\ \hline
 		Tamm & SCM & \red{nondissipative} HID & -- & $0$ &  --\\ 
 		& PMLID material & \red{nondissipative} HID & \tref{tab:qPMLID-HIDlless}& $2$ &  1 $p$- and 1 $s$-polarized\\ \hline
 		Dyakonov--Tamm & SCM & PMLID material & -- & $1$ &  No polarization state can be assigned\\ \hline
 	\end{tabular}
 \end{table*}

\subsection{Planar SCM/metal interface}
\label{sec:SCMmet}

 Two SPP-wave modes can be guided by the interface of the chosen SCM and metal, when $\psi=30$~deg. According to \tref{tab:qSCMmet},
one of these two modes will
propagate five times farther than the other mode, with the former possessing a higher phase speed
than the latter.

\begin{table}[h!]
\centering
\caption{\bf $\tq$, $\vph$, and  $\propdist$ for SPP waves guided by the SCM/metal interface
alone for  $\psi=30$~deg.}\vspace{4mm}
\begin{tabular}{ccccr}
\hline
Solution & Re$\tond{\tq}$ & Im$\tond{\tq}$ &  {$\vph/c\ped0$} & {$\propdist$ 
($\mu$m)}
 \\
\hline
$1$ & $1.18553$ & $3.48\times10^{-3}$ & $0.84350$   &$28.95$ \\
$2$ & $1.85217$ & $1.77\times10^{-2}$ & $0.53991$  &$5.69$ \\
\hline
\end{tabular}
  \label{tab:qSCMmet}
\end{table}

\subsection{Planar SCM/dissipative-HID interface}
\label{sec:lossyDT}
 The planar interface of the chosen SCM and the \red{nonplasmonic} but dissipative analog
of silver (with
$\eps_{dHID}=-\eps_{Ag}^\ast$)
by itself was found to guide two different Uller--Zenneck
waves (also classifiable as  Dyakonov--Tamm waves \cite{LP2007})
when $\psi=30$~deg; see Table~\ref{tab:qSCM-HID} for relevant data.

The substantial difference between the SPP waves in Sec.~\ref{sec:SCMmet}
and the Uller--Zenneck waves in this section is in the much shorter
propagation distances of the latter than of the former. This difference must be due
to the considerably smaller skin depth of silver ($\delta=26.5$~nm)
than of its \red{nonplasmonic} but dissipative analog ($\delta=642$~nm).
Another difference between the SPP and Uller--Zenneck waves 
is that, whereas the higher-$\vph$ SPP wave has the longer propagation
distance, the higher-$\vph$ Uller--Zenneck wave has the shorter propagation
distance.

\begin{table}[h!]
	\centering
\caption{\bf $\tq$, $\vph$, and  $\propdist$ for Uller--Zenneck (or Dyakonov--Tamm) waves guided by the SCM/dissipative-HID interface
alone for  $\psi=30$~deg.}\vspace{4mm}
\begin{tabular}{ccccr}
		\hline
Solution & Re$\tond{\tq}$ & Im$\tond{\tq}$ &  {$\vph/c\ped0$} & {$\propdist$ 
($\mu$m)}\\
		\hline
		$1$ & $1.06817$ & $5.90\times10^{-2}$ & $0.93618$ &  $1.71$ \\
		$2$ & $1.52148$ & $5.58\times10^{-2}$ & $0.65725$ & $1.81$ \\
		\hline
	\end{tabular}
	\label{tab:qSCM-HID}
\end{table}

\subsection{Planar SCM/\red{nondissipative}-HID interface}
\label{sec:DT} 

The planar interface of the chosen SCM and glass by itself was not found to guide  any ESW.
This is not surprising because the range of the refractive index of the lossless HID
material for the existence of a Dyakonov--Tamm wave is generally very small~\cite{LP2007}, and we conclude that
the chosen refractive index ($=1.6$) of glass must  lie outside that range.
 

\subsection{Planar PMLID/metal interface}
\label{sec:PMLIDmet}

 The planar interface of the chosen PMLID material and silver by itself was found to guide four  $p$- and three $s$-polarized  SPP
waves, with no dependence \red{on} $\psi$.  
These solutions are labeled as $p_{1,2,3,4}$ and $s_{1,2,3}$ in \tref{tab:qPMLIDmet}.
\red{An increase} in the number of SPP waves when the SCM is replaced by a PMLID material has been noticed
earlier \cite{PML-SCMmet,Atalla}. A high-phase-speed solution  (Re$\graff{\tq}<1$) \cite{Atalla}
exists for each polarization state.

\begin{table}[h!]
	\centering
\caption{\bf $\tq$, $\vph$, and  $\propdist$ for $p$- and $s$-polarized SPP waves guided by the PMLID/metal interface
alone. Solutions with  $\vph>c\ped0$ are
highlighted in bold font.}\vspace{4mm}
\begin{tabular}{ccccr}
		\hline
Solution & Re$\tond{\tq}$ & Im$\tond{\tq}$ &  {$\vph/c\ped0$} & {$\propdist$ 
($\mu$m)}\\
		\hline
		${\bf p_1}$ & ${\bf 0.93964}$ & ${\bf 5.21\times10^{-4}}$ & ${\bf1.06424}$ & ${\bf 193.37}$ \\
		$p_2$ & $1.36469$ & $1.56\times10^{-4}$ & $0.73277$ & $645.80$ \\
		$p_3$ & $1.59093$ & $1.49\times10^{-3}$ & $0.62856$ & $67.61$ \\
		$p_4$ & $2.27526$ & $3.81\times10^{-2}$ & $0.43951$ & $2.64$ \\
		${\bf s_1}$ & ${\bf 0.95783}$ & ${\bf 1.26\times10^{-3}}$ &  ${\bf 1.04403}$ & ${\bf 79.96}$ \\
		$s_2$ & $1.40658$ & $8.45\times10^{-4}$ & $0.71094$ & $119.22$ \\
		$s_3$ & $1.66592$ & $7.57\times10^{-4}$ & $0.60027$ & $133.08$ \\
		\hline
	\end{tabular}
	\label{tab:qPMLIDmet}
\end{table}

\subsection{Planar PMLID/dissipative-HID interface}
\label{sec:lossyT} 
The planar interface of the chosen PMLID material and the \red{nonplasmonic} but dissipative HID \red{material}
by itself was found to guide two different Uller--Zenneck
waves (also classifiable as  Tamm waves \cite{LP2007})
indepenently of $\psi$; see Table~\ref{tab:qPMLID-HID} for relevant data.
The substantial difference between the SPP waves in Table~\ref{tab:qPMLIDmet}
and the Uller--Zenneck waves in Table~\ref{tab:qPMLID-HID} is in the generally much shorter
propagation distances of the latter than of the former.  Again, this difference  must be ascribed
to the considerably smaller skin depth of silver ($\delta=26.5$~nm)
than of its \red{nonplasmonic} but dissipative analog ($\delta=642$~nm).

\begin{table}[h!]
	\centering
\caption{\bf $\tq$, $\vph$, and  $\propdist$ for $p$- and $s$-polarized Uller--Zenneck
(or Tamm) waves guided by the PMLID/dissipative-HID interface
alone. Solutions with  $\vph>c\ped0$ are
highlighted in bold font.}\vspace{4mm}
\begin{tabular}{ccccc}
		\hline
Solution & Re$\tond{\tq}$ & Im$\tond{\tq}$ &  {$\vph/c\ped0$} & {$\propdist$ 
($\mu$m)}\\
		\hline
		${\bf p_1}$ & ${\bf 0.92055}$ & ${\bf 3.96\times10^{-2}}$ &  ${\bf 1.08631}$ & ${\bf 2.54}$ \\
		$p_2$ & $1.52861$ & $5.52\times10^{-2}$ & $0.65419$ & $1.83$ \\
		$p_3$ & $1.71776$ & $1.46\times10^{-1}$ & $0.58215$ & $0.69$ \\
		${\bf s_1}$ & ${\bf 0.91802}$ & ${\bf 6.05\times10^{-2}}$ & ${\bf 1.08930}$ &  ${\bf 1.67}$ \\
		$s_2$ & $1.38091$ & $3.14\times10^{-2}$ & $0.72416$ & $3.21$ \\
		$s_3$ & $1.64373$ & $2.68\times10^{-2}$ & $0.60837$ & $3.76$ \\
		\hline
	\end{tabular}
	\label{tab:qPMLID-HID}
\end{table}

\subsection{Planar PMLID/\red{nondissipative}-HID interface}
\label{sec:T}
The planar interface of the chosen PMLID material and glass by itself was  found to guide  one $p$- and one $s$-polarized Tamm waves, without any dependence on $\psi$; see  \tref{tab:qPMLID-HIDlless} for data. Both partnering materials being \red{nondissipative}, 
the wavenumbers are  purely real, which means that both Tamm waves
 can be expected to propagate for very long distances by experimentalists. 

\begin{table}[h!]
	\centering
\caption{\bf $\tq$ and $\vph$  for $p$- and $s$-polarized Tamm
waves guided by the PMLID/glass interface
alone.  }\vspace{4mm}
	\begin{tabular}{ccc}
		\hline
		Solution & $\tq$  &  {$\vph/c\ped0$}
\\
		\hline
		$p_1$ & $1.72282$ & $ 0.58044$ \\
		$s_1$ & $1.74660$ & $0.57254$\\
		\hline
	\end{tabular}
	\label{tab:qPMLID-HIDlless}
\end{table}

\subsection{Planar SCM/PMLID interface }
\label{sec:DTd=0}

\begin{figure}[h!]
\centering
\includegraphics[width=0.8\linewidth]{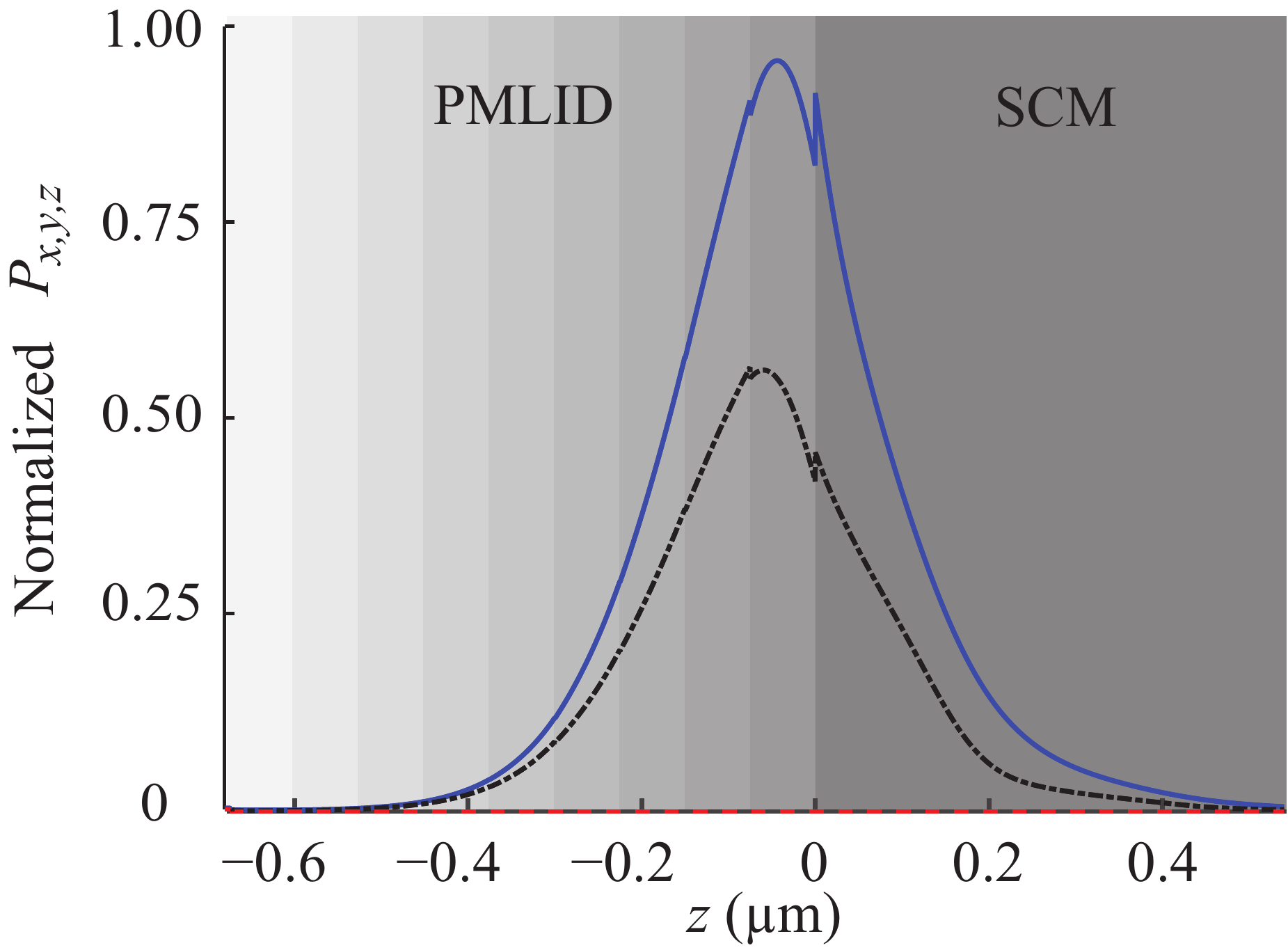}
\caption{Spatial variations of $P_x(0,0,z)$ (blue solid lines), $P_y(0,0,z)$ (black dashed-dotted lines) and $P_z(0,0,z)$ (red dashed lines) with respect to $z$	 of
the Dyakonov--Tamm wave guided by the  planar SCM/PMLID interface when $\psi=30^\circ$.}
\label{fig:DTd=0_psi30_Poynt}
\end{figure}

Only one ESW was found to be guided by the SCM/PMLID interface when $\psi=30$~deg. Classified as a Dyakonov--Tamm wave, its normalized wavenumber $\tq=1.77612$ and its phase speed $\vph=0.56302c\ped0$.
  Spatial variations of the Cartesian components  of its	time-averaged Poynting vector  
  $\#P(0,0,z)$ are presented in \fref{fig:DTd=0_psi30_Poynt}.

\section{CGWs guided by {\sf A}/{\sf B}/{\sf C} structure}
\label{sec:twointerfaces}
Now let us turn to  the problem schematically illustrated in \fref{fig:schematic},
with  materials {\sf A} and {\sf C} being the chosen PMLID material and SCM, respectively, while material {\sf B} is isotropic and homogeneous. The thickness $L$ of the sandwiched layer was varied from $L_{min}=0$ to $L_{max}=120$~nm. With some degree
of approximation, this problem can be practically implemented with
the  {\sf A}/{\sf B}/{\sf C}
structure of sufficiently large thickness and finite
width interposed between two waveguide sections.

\subsection{Planar PMLID/metal/SCM structure}\label{sec:PMLIDmetSCM} 

Let us begin by choosing material $\sf B$ to be silver (i.e., $\eps_{\sf B}=\eps_{Ag}$). Since materials $\sf A$ and $\sf C$
are dielectric, the CGWs guided by this structure must be classified as compounded from
SPP waves  for $L>0$. However, this compounding is
more complicated than in two predecessor papers wherein either $\sf A$
\cite{NoiOC15} or $\sf C$ \cite{NoiJNP15}
was isotropic and homogeneous.
The number of CGWs and their wavenumbers   depend on $\psi$ due to the anisotropy of the SCM. Except in special cases, no  polarization state can be assigned to any of the CGWs. 

\begin{figure}[h!]
	\centering
	\includegraphics[width=0.8\linewidth]{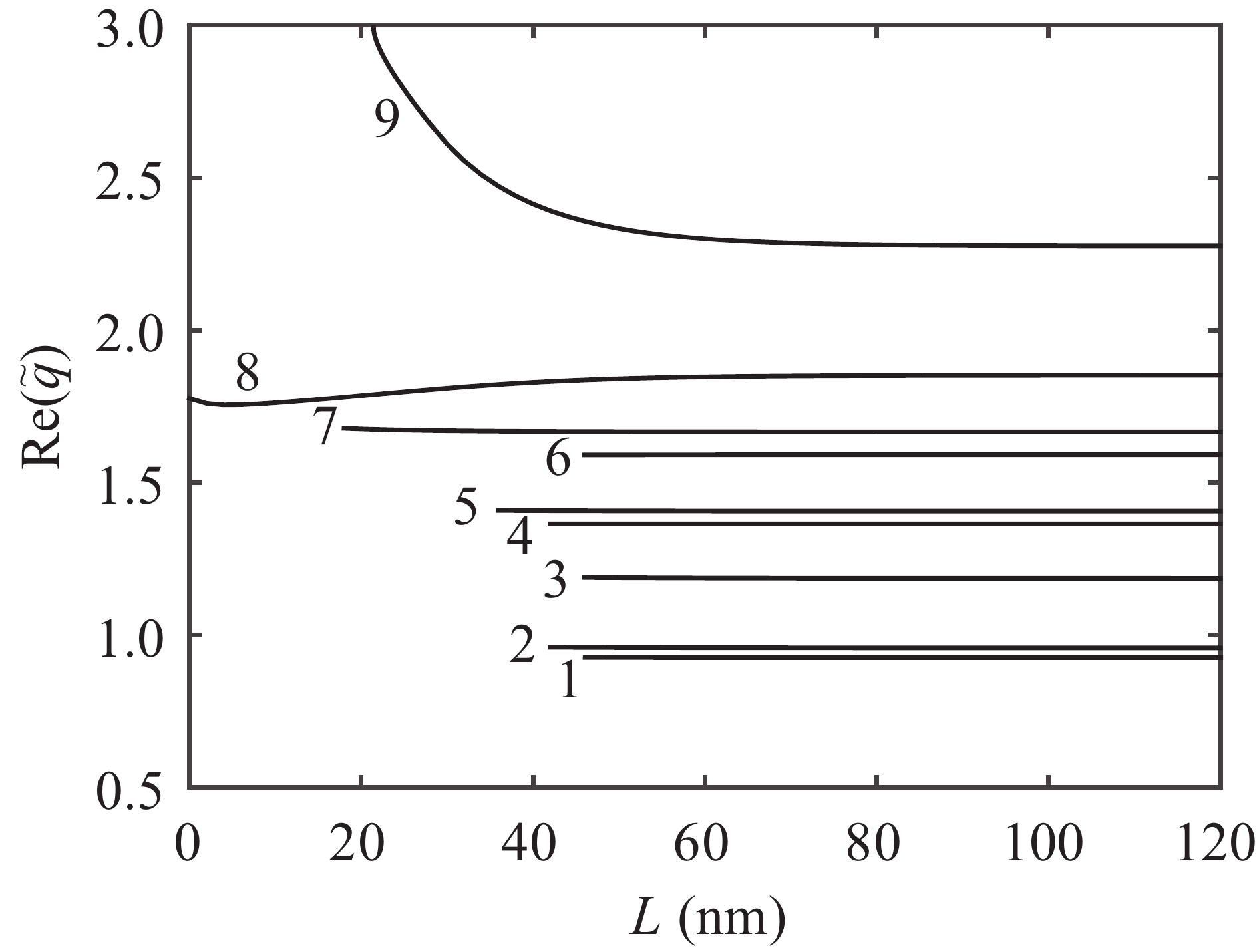}
	\caption{Variation of ${\rm Re}(\tq)$ with  thickness $L$  of the metal layer for  the
	CGWs guided by PMLID/metal/SCM structure when  $\psi=30$~deg.
	\label{fig:ReqLm_psi30}}
\end{figure}

\begin{figure}[h!]
	\centering
	\includegraphics[width=0.8\linewidth]{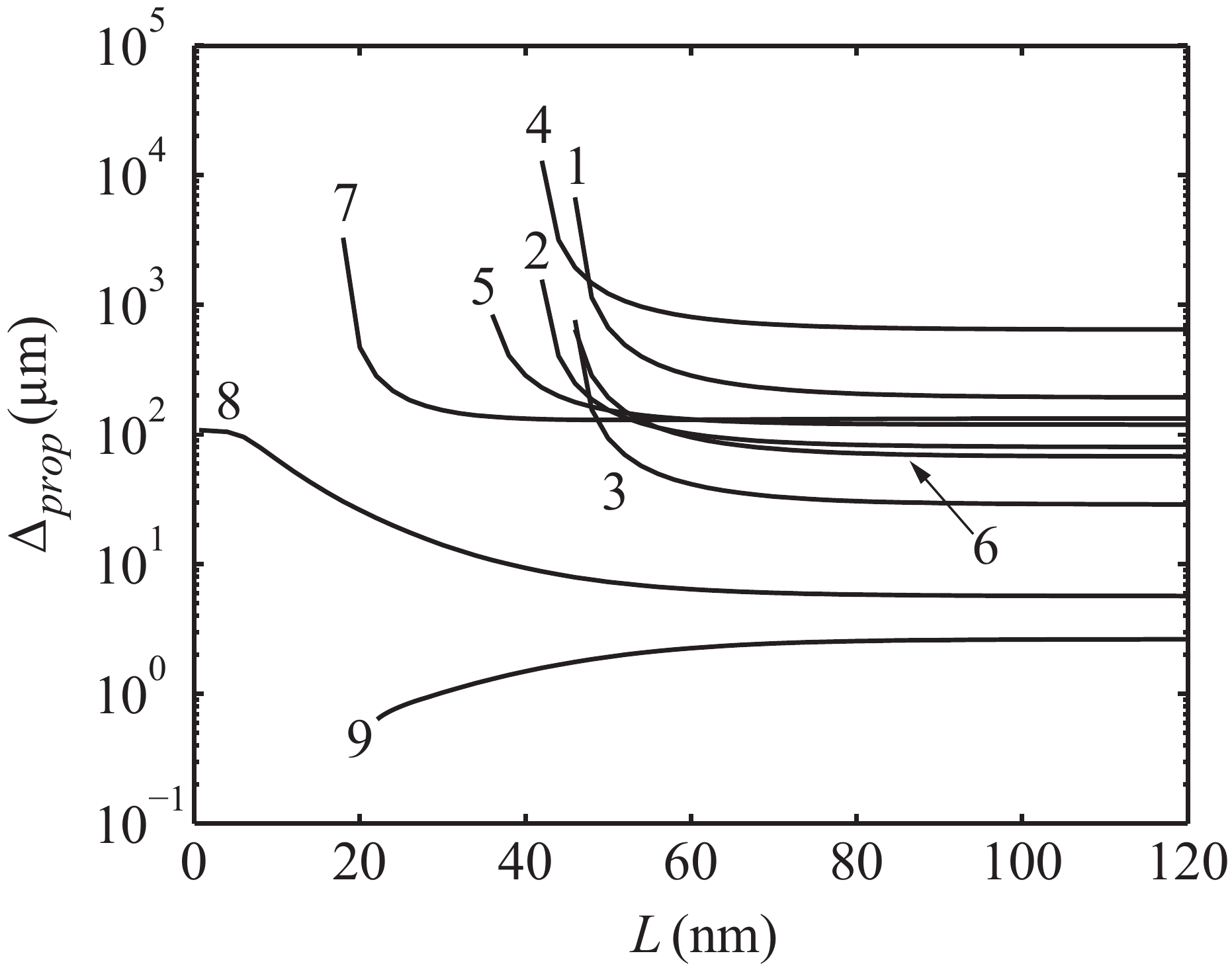}
	\caption{Variation of $\propdist$ with  thickness $L$  of the metal layer for  the
	CGWs guided by PMLID/metal/SCM structure when  $\psi=30$~deg.}
	\label{fig:DpropLm_psi30}
\end{figure}

Depending on the value of $L\in(0,120]$~nm, as many as nine different CGWs can be guided by the structure.  Figures~\ref{fig:ReqLm_psi30} and
\ref{fig:DpropLm_psi30}, respectively, show ${\rm Re}(\tq)$ and  $\propdist$ 
as functions of $L$, the wavenumbers $q$ being organized into nine different branches.
Solutions on branches numbered $1$ and $2$ in the figures have phase speeds exceeding
$c\ped0$, \red{whereas} the other seven branches have solutions with phase speeds below
$c\ped0$.

For metal  thickness $L$ significantly greater than the skin depth $\delta=26.5$~nm of silver,  $ \tq $ and $\propdist$ assume  steady values. Their values for the thickest metal layer considered ($L=120$ nm)   are reported in \tref{tab:qLm_psi30}. In that table, solutions 1, 4, 6, and 9 are practically the same   as  for the $p$-polarized SPP waves guided by the 
PMLID/metal interface alone (see \tref{tab:qPMLIDmet}); solutions 2, 5, and 7 are almost the same   as for the 
$s$-polarized SPP waves guided by the PMLID/metal interface
alone (see \tref{tab:qPMLIDmet}); and solutions 3 and 8  are almost the same as  for the SPP waves guided by planar metal/SCM interface alone (see \tref{tab:qSCMmet}). In other words,
the PMLID/metal and metal/SCM interfaces in the PMLID/metal/SCM structure are virtually decoupled from each other for $L\gg\delta$. 

\begin{figure}[h!]
	\begin{center}
			\includegraphics[width=0.8\linewidth]{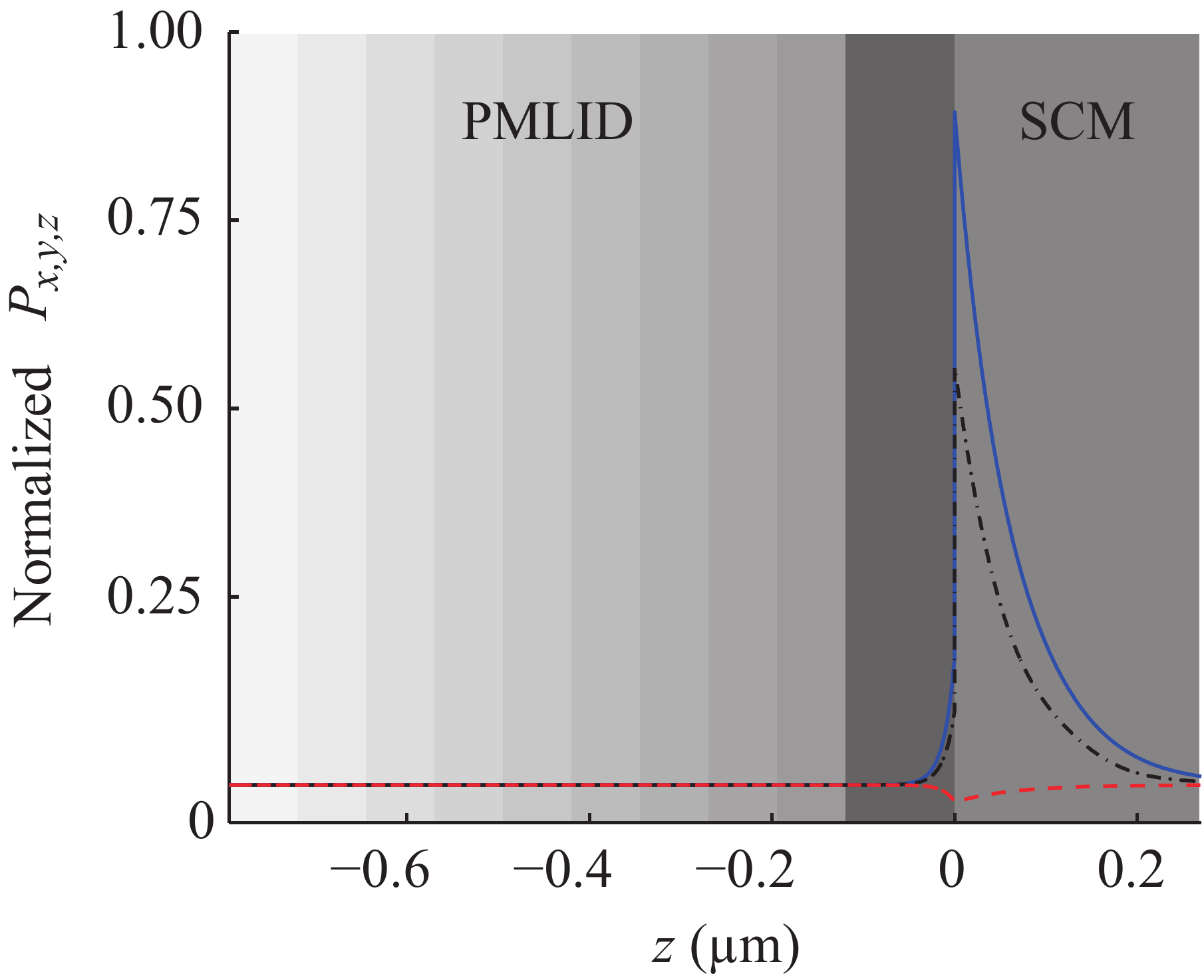}
	\end{center}
\caption{Spatial variations of $P_x(0,0,z)$ (blue solid lines), $P_y(0,0,z)$ (black dashed-dotted lines) and $P_z(0,0,z)$ (red dashed lines) with respect to $z$	for solution 8 of \tref{tab:qLm_psi30}. The CGW is guided by the PMLID/metal/SCM structure, when $L= 120$ nm and $\psi=30$~deg.
\label{fig:q8met}}
\end{figure}

\begin{figure}[!h]
	\begin{center}
			\includegraphics[width=0.8\linewidth]{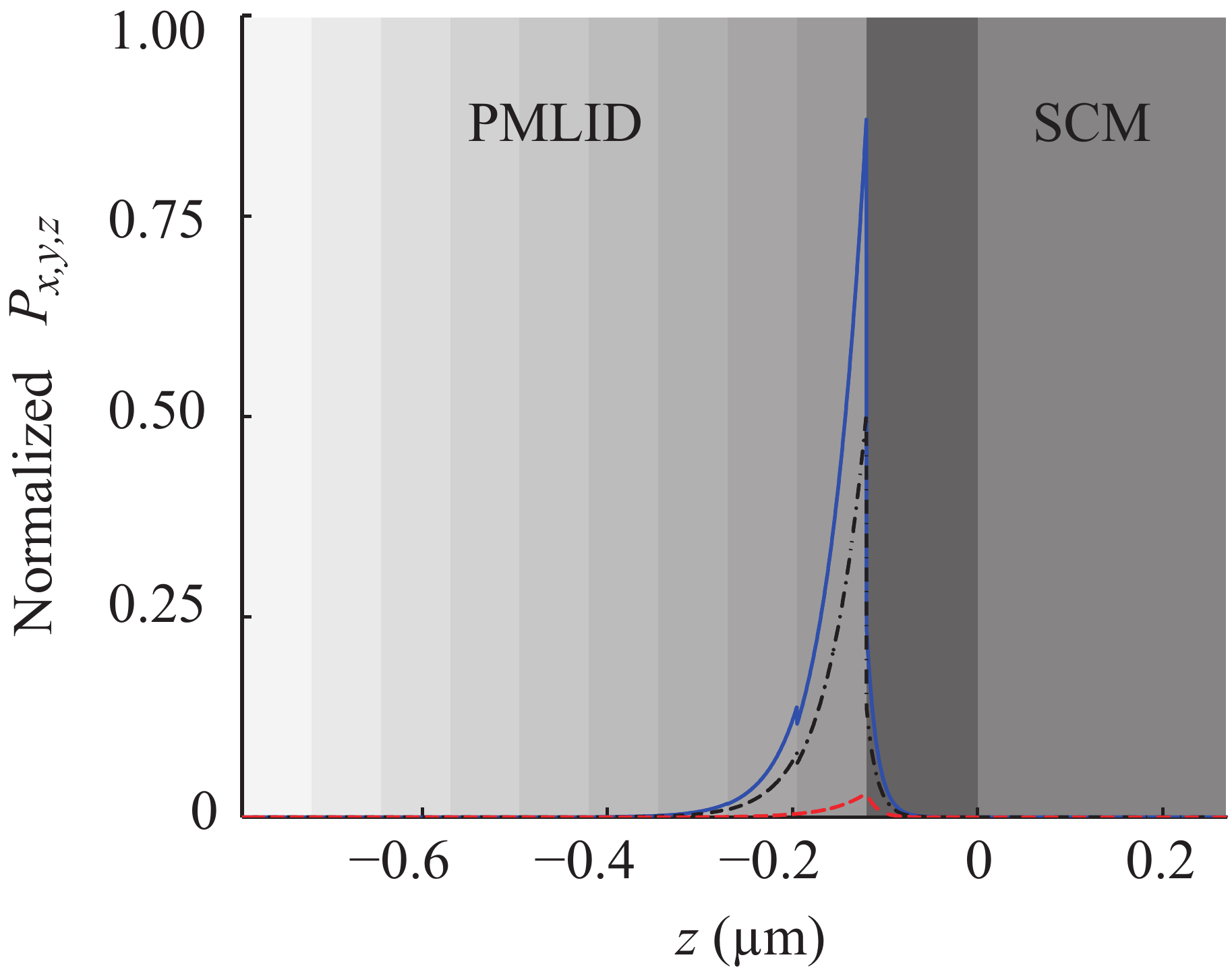}
	\end{center}
\caption{Same as Fig.~\ref{fig:q8met}, but for  solution 9 of \tref{tab:qLm_psi30}.
\label{fig:q9met}}
\end{figure}

This decoupling is evident in the spatial distributions of $\#P(0,0,z)$ plotted in Figs.~\ref{fig:q8met}, and \ref{fig:q9met}
for solutions 8, and 9 for $L= 120$~nm and $\psi=30$~deg. The CGWs are  bound to either the  PMLID/metal interface or the metal/SCM interface, but not to both.  

\begin{table}[h!]\setlength\tabcolsep{4pt}
	\centering
	\caption{\bf $\tq$, $\vph$, and  $\propdist$ for  CGWs guided by the PMLID/metal/SCM structure when $L= 120$ nm and $\psi=30$~deg. Solutions with $\vph$ exceeding $c\ped0$ are
highlighted in bold font. Threshold value $L_{th}$ of the metal thickness  below which a solution branch ceases to exist is also reported (correct to the nearest nm), except for solution branch 9.}\vspace{4mm}
	\begin{tabular}{ccccrc}
		\hline
		Soln.  & Re$\graff{\tq}$ & Im$\graff{\tq}$ & $\vph/c\ped0$ &$\propdist$ ($\mu$m) & $L_{th}$ (nm) \\
		\hline
		1 &  ${\bf 0.93964}$ & ${\bf 5.19\times 10^{-4}}$ & ${\bf 1.06424}$ & ${\bf 194.11}$  & ${\bf 46}$ \\
		2 & ${\bf 0.95784}$ & ${\bf 1.25\times 10^{-3}}$ & ${\bf 1.04402}$ & ${\bf 80.60}$  & ${\bf 42}$ \\
		3 & $1.18554$ & $3.48\times 10^{-3}$ & $0.84350$ & $28.95$  & $46$ \\
		4 & $1.36469$ & $1.56\times 10^{-4}$ & $0.73277$ & $645.80$  & $42$\\
		5 & $1.40658$ & $8.44\times 10^{-4}$ & $0.71094$ & $119.37$ & $36$ \\
		6 & $1.59093$ & $1.48\times 10^{-3}$ & $0.62856$ & $68.07$ & $46$ \\
		7 & $1.66592$ & $7.57\times 10^{-4}$ & $0.60027$ & $133.08$ & $18$ \\
		8 & $1.85214$ & $1.77\times 10^{-2}$ & $0.53992$ & $5.69$ & $0$ \\
		9 & $2.27538$ & $3.81\times 10^{-2}$ & $0.43949$ & $2.64$ & $<21$ \\
		\hline
	\end{tabular}
	\label{tab:qLm_psi30}
\end{table}

Coupling of the PMLID/metal and metal/SCM interfaces  enhances as $L$ is diminished.
Each solution branch in Figs.~\ref{fig:ReqLm_psi30} and \ref{fig:DpropLm_psi30} exists
only for $L > L_{th}$,  the threshold value  $L_{th}$ for each  branch (except branch 9) being reported in the last column of \tref{tab:qLm_psi30}. We were unable to determine $L_{th}$ for solution branch 9, because ${\rm Re}(\tq)$ exceeded $3$ as $L$ decreased below $21$~nm and thus fell out of the range
$0<{\rm Re}(\tq) \leq 3$.

Solution branches 1 to 6 have  $L_{th}>\delta=26.5$~nm, implying relatively  weak coupling between the
PMLID/metal and metal/SCM interfaces for $L>L_{th}$. Still,  Figs.~\ref{fig:ReqLm_psi30} and \ref{fig:DpropLm_psi30} indicate that $\tq$ is substantially impacted by the coupling for $L\in[L_{th},2\delta]$.

\begin{figure}[h!]
	\begin{center}
			\includegraphics[width=0.8\linewidth]{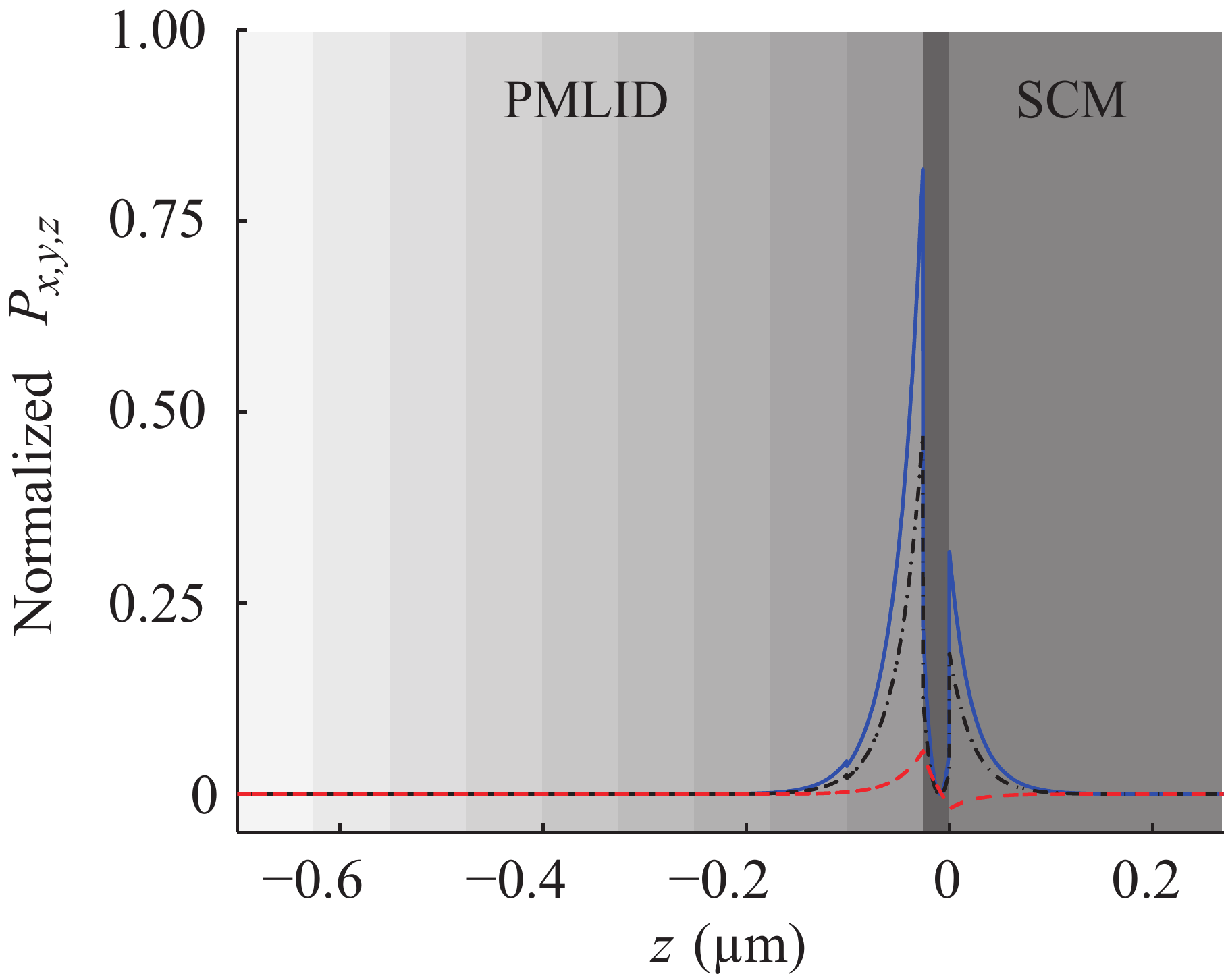}
	\end{center}
\caption{Spatial variations of $P_x(0,0,z)$ (blue solid lines), $P_y(0,0,z)$ (black dashed-dotted lines) and $P_z(0,0,z)$ (red dashed lines) with respect to $z$	for solution branch 9 for the PMLID/metal/SCM structure, when $L= 26$~nm and $\psi=30$~deg.
\label{fig:q9met_strong}}
\end{figure}

Solution branch 9  exists for $L\approx \delta$,
indicating stronger coupling between the two interfaces. This is confirmed by the plots of $\#P(0,0,z)$ versus $z$ for $L=26$~nm in
Fig.~\ref{fig:q9met_strong}.  Clearly, both the PMLID material and the SCM
carry the energy of the CGW, the former more than the latter. Thus, the characteristics
of SPP waves of two different types have been mixed in the CGW.  SPP waves of one type
are guided by the PMLID/metal interface and have specific polarization states,
whereas no polarization state can be assigned for SPP waves of the other type 
as they are guided by the metal/SCM interface \cite{AkhBook,PML-SCMmet}.
Similar conclusions can be
drawn for solution branch 7 as well, for which $L_{th}\approx0.68\delta$.

\begin{figure}[h!]
	\begin{center}
			\includegraphics[width=0.8\linewidth]{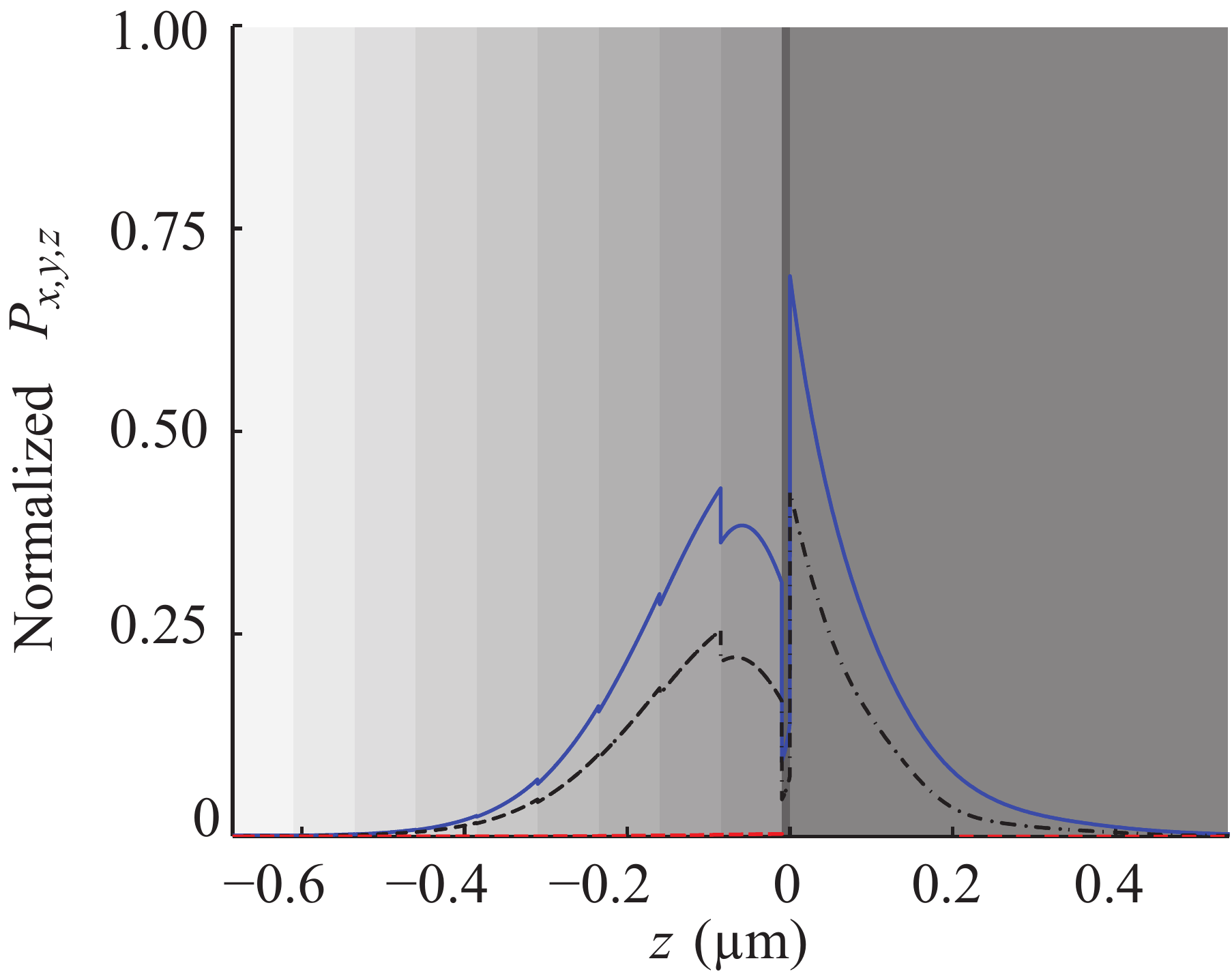}
	\end{center}
	\caption{Spatial variations of $P_x(0,0,z)$ (blue solid lines), $P_y(0,0,z)$ (black dashed-dotted lines) and $P_z(0,0,z)$ (red dashed lines) with respect to $z$	for solution branch 8 for the PMLID/metal/SCM structure, when $L= 10$~nm and $\psi=30$~deg.
\label{fig:setSCM@Lth}}
\end{figure}

Finally,  solution branch 8 has $L_{th}=0$. Hence, solutions must exist even if $L \ll \delta$,
which is indicative of very strong coupling between the PMLID/metal and metal/SCM interfaces. The strong coupling is very clear in the
spatial profile of $P(0,0,z)$ presented in Fig.~\ref{fig:setSCM@Lth} for $L=10$~nm, with the energy of the CGW carried almost equally by the PMLID material and the SCM.

Furthermore, when $L=0$, the PMLID/metal/SCM structure collapses into the PMLID/SCM structure, $\tq$ on the solution branch 8 in Figs.~\ref{fig:ReqLm_psi30} and \ref{fig:DpropLm_psi30} acquires the purely real value $1.77612$,   the spatial variations of the Cartesian components  of the	time-averaged Poynting vector  
  $\#P(0,0,z)$ of the CGW are the same as presented in \fref{fig:DTd=0_psi30_Poynt}.
 and the CGW converts into the Dyakonov--Tamm wave discussed in
 \red{Sec.~\ref{sec:DTd=0}}.  

\subsection{Planar PMLID/dissipative-HID/SCM structure} \label{sec:UllZenn}

When material {\sf B} is the \red{nonplasmonic} analog of silver (i.e., $\eps_{\sf B}=\eps_{dHID}=\eps_{Ag}^\ast$), as many as five different CGWs can be guided by 
the PMLID/dissipative-HID/SCM structure for $L\in[0,120]$~nm and
 $\psi=30$~deg. These CGWs for $L>0$ are compounded from Uller--Zenneck 
waves of Secs.~\ref{sec:DT} and  \ref{sec:T};
 alternatively, they are compounded from the Dyakonov--Tamm waves of Sec.~3.\ref{sec:DT}
 and the Tamm waves of Sec.~\ref{sec:T}.
 
\begin{figure}[h!]
	\centering
	\includegraphics[width=0.8\linewidth]{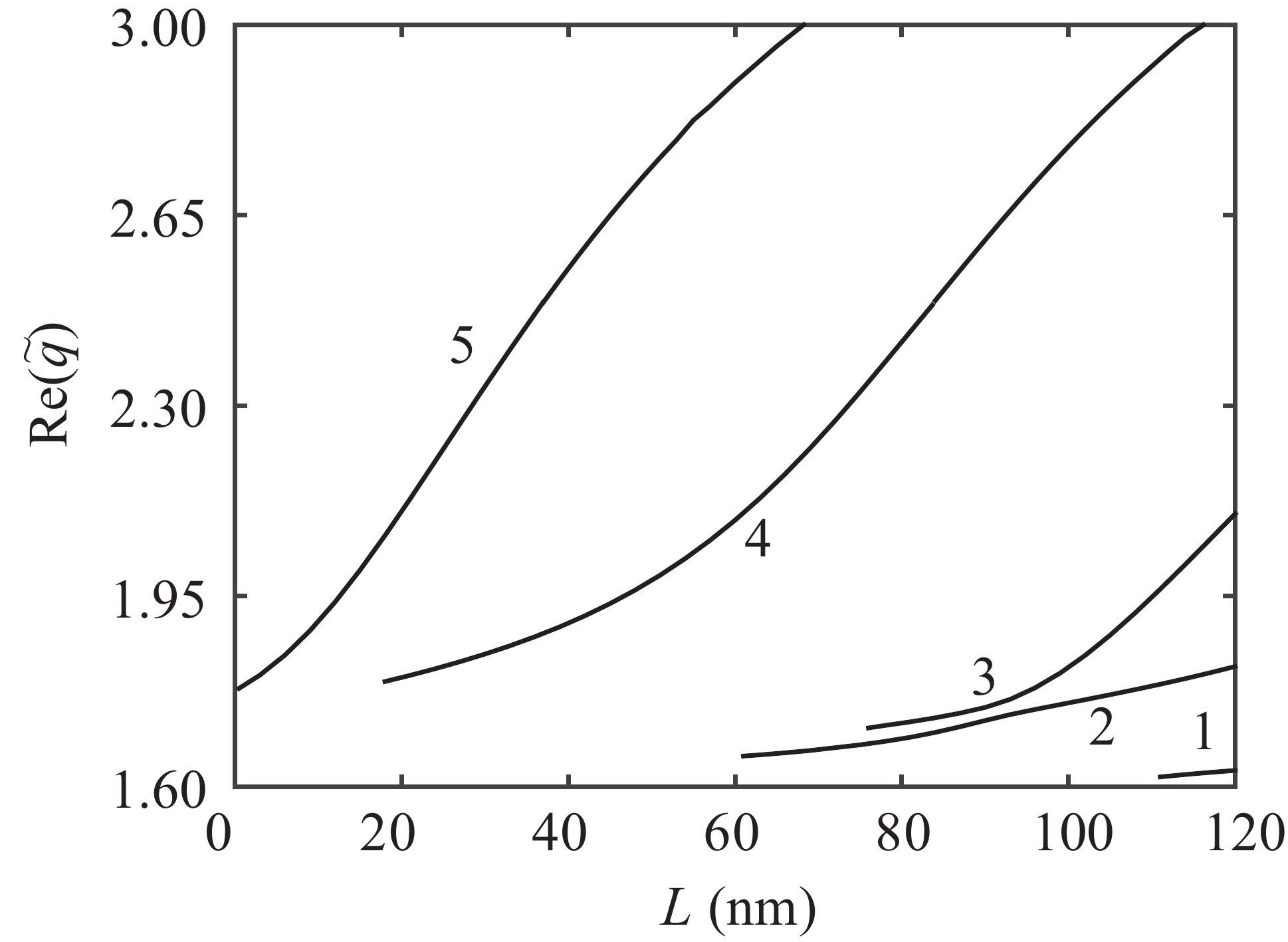}
	\caption{Variation of ${\rm Re}(\tq)$ with  thickness $L$  of the dissipative-HID layer for  the
	CGWs guided by PMLID/dissipative-HID/SCM structure   when  $\psi=30$~deg.
	\label{fig:ReqLd_psi30}}
\end{figure}

\begin{figure}[h!]
	\centering
	\includegraphics[width=0.8\linewidth]{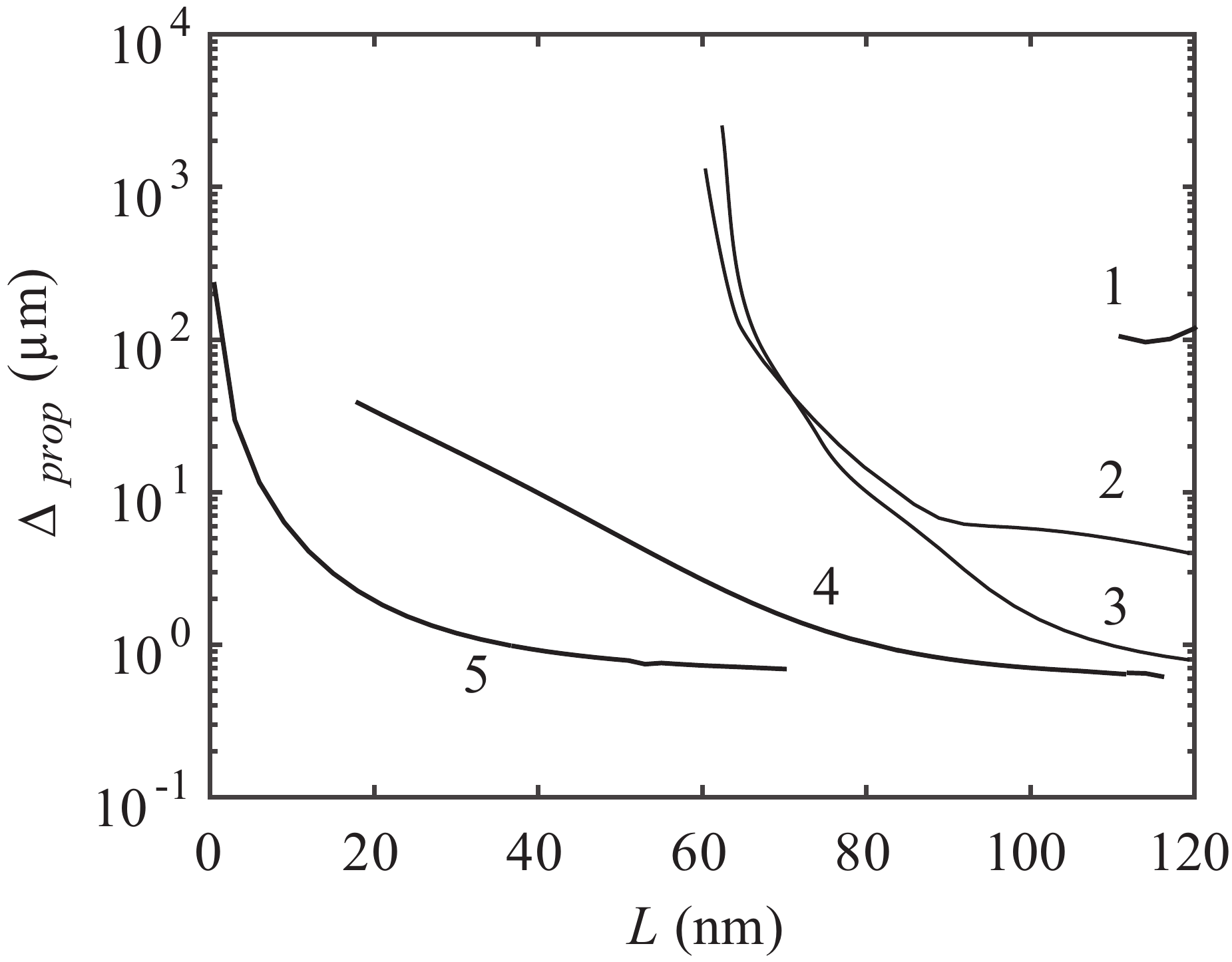}
	\caption{Variation of $\propdist$ with  thickness $L$  of the metal layer for  the
	CGWs guided by PMLID/dissipative-HID/SCM structure   when  $\psi=30$~deg.
	\label{fig:DpropLd_psi30}}
\end{figure}

Figures~\ref{fig:ReqLd_psi30} and \ref{fig:DpropLd_psi30} show Re$\tond{\tq}$ and $\propdist$ as   functions of $L\in(0,120]$~nm, respectively. The  solutions $q$ are organized into five different branches. On every branch, both $\vph$ and $\propdist$ decrease as $L$ increases. No solutions with phase speed exceeding $c\ped0$ were found, most likely because (i) although the PMLID/dissipative-HID interface alone does support
high-phase-speed Uller--Zenneck (or Tamm)    waves (see \tref{tab:qPMLID-HID}),
the SCM/dissipative-HID interface alone does not support
high-phase-speed Uller--Zenneck (or Dyakonov--Tamm)    waves (see \tref{tab:qSCM-HID},
and (ii) $L_{max}$ is much smaller than the skin depth
$\delta=642$~nm in $\sf B$.

Values of $\tq$ and $\propdist$ when  $L=3000$~nm   are reported in \tref{tab:qLd_psi30}. In this table, solutions $b$ and $h$ have values very close to those for  two $p$-polarized Uller--Zenneck waves guided by the 
PMLID/dissipative-HID interface alone (see \tref{tab:qPMLID-HID}); solutions $a$, $d$, and $g$ have values very close to those for the 
$s$--polarized Uller--Zenneck waves guided by the PMLID/dissipative-HID interface
alone (see \tref{tab:qPMLID-HID}); and solution $c$ has a value very close to that for an Uller--Zenneck wave guided by planar SCM/dissipative-HID interface alone (see \tref{tab:qSCM-HID}). Solutions $e$ and $f$ in \tref{tab:qLd_psi30} are \red{somewhat, but 
not
very,} close to a solution
in  \tref{tab:qSCM-HID} and a solution in \tref{tab:qPMLID-HID}, indicating that the two
interfaces are coupled even for $L\simeq4.67\delta$. Their decoupling would require even larger values of the ratio $L/\delta$.

\begin{table}[h!]\setlength\tabcolsep{6pt}
	\centering
	\caption{\bf $\tq$, $\vph$, and  $\propdist$ for  CGWs guided by the PMLID/dissipative-HID/SCM structure when   $L=3000$~nm and $\psi=30$~deg. Solutions with $\vph$ exceeding $c\ped0$ are highlighted in bold font.}\vspace{4mm}
		\begin{tabular}{ccccr}
		\hline
		Soln.  & Re$\graff{\tq}$ & Im$\graff{\tq}$ & $\vph/c\ped0$ &$\propdist$ ($\mu$m) \\
		\hline
		$a$ & ${\bf 0.91803}$ & ${\bf 6.06\times 10^{-2}}$ & ${\bf 1.08929}$ & ${\bf 1.66}$  \\
		$b$ & ${\bf 0.92057}$ & ${\bf 3.97\times 10^{-2}}$ & ${\bf 1.08663} $ & ${\bf 2.54}$  \\
		$c$ & $1.06833$ & $5.90\times 10^{-2}$ & $0.93604 $ & $1.71$  \\
		$d$ & $1.38090$ & $3.14\times 10^{-2}$ & $0.72417 $ & $3.21$  \\
		$e$ & $1.51980$ & $5.85\times 10^{-2}$ & $0.65798 $ & $1.72$ \\
		$f$ & $1.52934$ & $5.27\times 10^{-2}$ & $0.65388 $ & $1.91$ \\
		$g$ & $1.64373$ & $2.68\times 10^{-2}$ & $0.60837 $ & $3.76$ \\
		$h$ & $1.71959$ & $1.44\times 10^{-1}$ & $0.58153 $ & $0.70$ \\
		\hline
	\end{tabular}
	\label{tab:qLd_psi30}
\end{table}

Whereas the number of solution branches is five in Figs.~\ref{fig:ReqLd_psi30}
and ~\ref{fig:DpropLd_psi30} for $L\leq 120$~nm, the number of solutions is eight
in \tref{tab:qLd_psi30} for $L=3000$~nm. Eight must be the largest number of solutions
possible, based on the data in Tables~\ref{tab:qSCM-HID} and \ref{tab:qPMLID-HID}.
Since each solution branch will exist only for $L>L_{th}$, three solution branches
must have $L_{th}\in(120,3000)$~nm.  

\begin{figure}[h!]
	\begin{center}
			\includegraphics[width=0.8\linewidth]{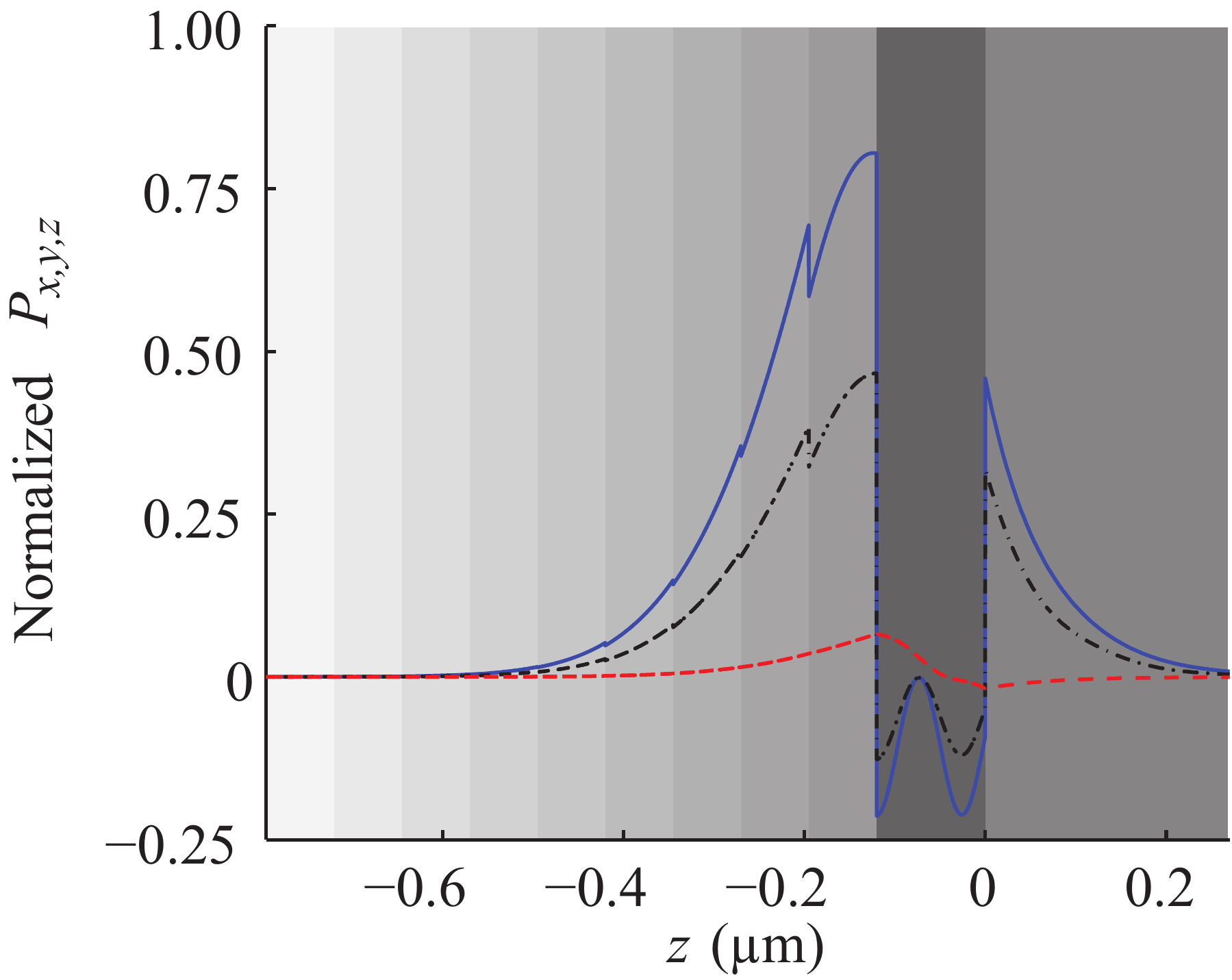}
	\end{center}
	\caption{Spatial variations of $P_x(0,0,z)$ (blue solid lines), $P_y(0,0,z)$ (black dashed-dotted lines) and $P_z(0,0,z)$ (red dashed lines) with respect to $z$	for solution 2 guided by the PMLID/dissipative-HID/SCM structure, when $L= 120$~nm and $\psi=30$~deg.}
	\label{fig:lossyHID_Poynt_Lmax}
\end{figure}

\fref{fig:lossyHID_Poynt_Lmax} present the
spatial profiles of $\#P\tond{0,0,z}$ for
solution 2 in \fref{tab:qLd_psi30} when  $L=120$~nm. A fraction of the energy of the CGW resides in both the PMLID material and the SCM, with the former carrying more energy than the latter. Clearly then, the characteristics
of Uller--Zenneck waves of two different types have been mixed in the CGW.  Uller--Zenneck waves of one type (alternatively, Tamm waves)
are guided by the PMLID/dissipative-HID interface and have specific polarization states,
whereas no polarization state can be assigned for Uller--Zenneck waves of the other type
(alternatively, Dyakonov--Tamm waves)  
as they are guided by the dissipative-HID/SCM interface \cite{LP2007,AkhBook}.
Similar conclusions hold for the other four solution branches in \fref{tab:qLd_psi30}.

Only one solution branch in Figs.~\ref{fig:ReqLd_psi30}
and ~\ref{fig:DpropLd_psi30}
has $L_{th}=0$. Solution 5 is identical
to that for the sole Dyakonov--Tamm wave guided by the PMLID/SCM interface by itself
in Sec.~\ref{sec:DTd=0}.

\subsection{Planar PMLID/\red{nondissipative}-HID/SCM structure}\label{sec:losslessUllZen}

When material {\sf B} is \red{nondissipative} (i.e., $\eps_{\sf B}=\eps_{glass}=2.56$), only one solution was found  in the range $L\in\quadr{0,120}$~nm. \fref{fig:losslessq(d,psi)} depicts $\tq$ as a function of $L$. Since all the materials involved are \red{nondissipative} $\tq$ is obviously real-valued so that, theoretically, $\propdist\rightarrow\infty$ and the CGW wave can propagate indefinitely. As discussed in Sec.~\ref{sec:T}, there will always be some dissipation so that $\propdist$ can be large but not infinite.

\begin{figure}[h!]
\centering
\includegraphics[width=0.8\linewidth]{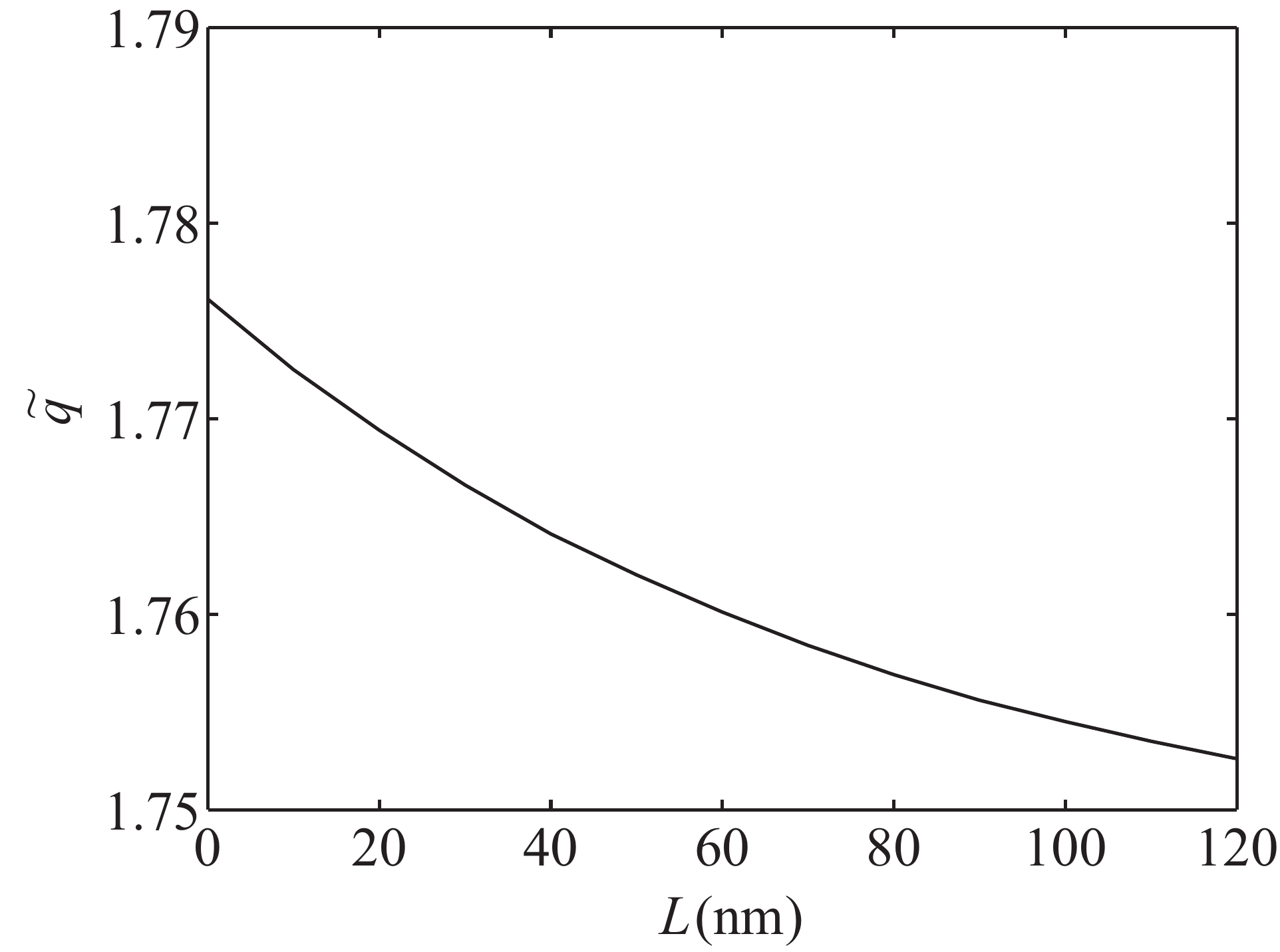}
\caption{Variation of $\tq$ with the thickness $L$ of the \red{nondissipative} HID layer for the CGW guided by the PMLID/\red{nondissipative}-HID/SCM structure when $\psi=30 \deg$.\label{fig:losslessq(d,psi)}}
\end{figure}

For a glass layer much thicker than $L_{max}=120$~nm (i.e., $L=1000$~nm),  two solutions were found. Their values, reported in \tref{tab:qHIDlless}, are the same as for the Tamm waves guided by the PMLID/glass interface alone (\tref{tab:qPMLID-HIDlless}) in Sec.~\ref{sec:T}. Solution 2 in \tref{tab:qHIDlless} is the asymptotic value of the curve  in \fref{fig:losslessq(d,psi)} as the thickness $L$ of the  glass layer increases, whereas
solution 1 has obviously a threshold value $L_{th}> 120$~nm.

\begin{table}[h!]
	\centering
	\caption{\bf $\tq$ and $\vph$ for  CGWs guided by the {PMLID/glass/} SCM structure, when   $L=1000$~nm and $\psi=30$~deg.}\vspace{4mm}
	\begin{tabular}{ccc}
		\hline
		Solution & $\tq$ & $\vph/c\ped0$\\
		\hline
		1 & $1.72282$ & $0.58044$\\
		2 & $1.74660$ & $0.57254$\\
		\hline
	\end{tabular}
	\label{tab:qHIDlless}
\end{table}

It is worth pointing out that the solutions found for the 
PMLID/\red{nondissipative}-HID/SCM structure for large $L$ are the
same as for the PMLID/\red{nondissipative}-HID interface alone.
Because the HID is \red{nondissipative} (i.e., the skin depth $\delta$ is infinite), this observation means that the   coupling between the two interfaces of the sandwiched layer is not only affected by the ratio $L/\delta$ but also by the  rates at which the fields
of the
 ESWs guided individually by the two interfaces decay away from the respective guiding interfaces.
 
\begin{figure}[h!]
	\begin{center}
		\includegraphics[width=0.8\linewidth]{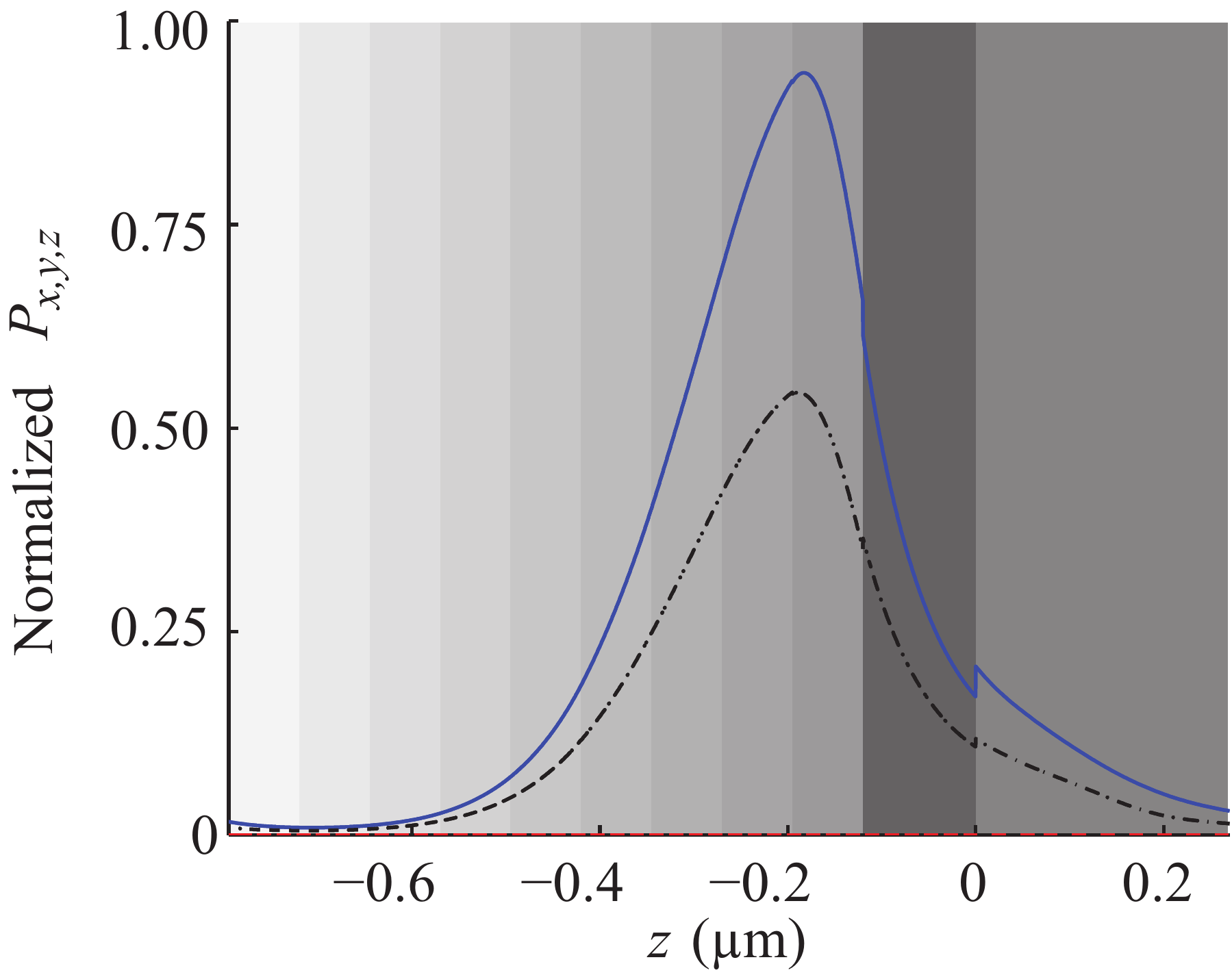}
	\end{center}
	\caption{Spatial variations of $P_x(0,0,z)$ (blue solid lines), $P_y(0,0,z)$ (black dashed-dotted lines) and $P_z(0,0,z)$ (red dashed lines) with respect to $z$ for the PMLID/glass/SCM structure, when $L= 120$~nm and $\psi=30$~deg.\label{fig:q1_Ld_120}}
\end{figure}

Even when the skin depth is infinite, a strong coupling between the 
PMLID/\red{nondissipative}-HID and the  \red{nondissipative}-HID/SCM 
interfaces is expected for $L\in\left(0,120\right]$~nm. This is confirmed by the plot of $\textbf{P}\tond{0,0,z}$ vs. $z$ for $L=120$~nm in \fref{fig:q1_Ld_120}. Clearly, both the PMLID and the SCM carry the energy of the CGW, the former more than the latter, and the characteristics
of Tamm and Dyakonov--Tamm waves  have been mixed in the CGW \cite{AkhBook,LP2007,YYH}.

\section{Concluding remarks}\label{sec:cr}

We solved the boundary-value problem for electromagnetic   waves guided by a layer of a homogeneous and isotropic (metal or dielectric) material sandwiched between a structurally chiral material   and a periodically multi-layered isotropic dielectric  material. Though this problem  is practically unimplementable \textit{sensu strictissimo} because it involves two half spaces, it provides information on the propagation  distance, otherwise unachievable. 
But it can be implemented in an approximate sense, with a PMLID/sandwiched-material/SCM structure of sufficiently large thickness and finite
width interposed between two waveguide sections.

We found that  compound guided waves can propagate bound to both interfaces of the sandwiched material with energy distributed in both the SCM and the PMLID materials,   if the sandwiched layer is sufficiently thin. Hence, CGWs that mix together the characteristics of SPP waves, Tamm waves,  Dyakonov--Tamm waves, and Uller--Zenneck waves are theoretically possible, exemplifying the inherent commonality between these electromagnetic surface waves. A multiplicity of CGWs can exist, depending on the thickness of the sandwiched layer. All of them differ in phase speed, attenuation rate, and field profile, even though all are excitable at the same frequency. For any thickness of the sandwiched layer, at least one CGW exists.

The coupling between the two  faces of the sandwiched layer, giving rise to CGWs, is affected by two distinct mechanisms: (i) the ratio of the thickness of the sandwiched layer to the skin depth in that material and (ii) the rates at which the fields of the ESWs guided individually by the two interfaces decay away from their respective guiding interfaces.

Finally,  CGWs have been found for an all-dielectric-configuration where the materials used can even be extremely weakly dissipative. Such a configuration could be very useful for sensing applications in special cases where the presence of a metal is unwanted, e.g. in an electrically hazardous environment.

\end{document}